\documentclass[
    ,final            
  ]
  {aipproc}

\layoutstyle{6x9}

\begin{document}

\title{Cluster Core Heating from Merging Subclusters}


\author{John ZuHone}{
  address={Harvard-Smithsonian Center for Astrophysics, 60 Garden St., Cambridge, MA 02138, USA}
}

\author{Maxim Markevitch}{
  address={Harvard-Smithsonian Center for Astrophysics, 60 Garden St., Cambridge, MA 02138, USA}
}

\begin{abstract}
Though feedback from central active galactic nuclei provides an attractive solution to the problem of overcooling in galaxy cluster cores, another possible source of heating may come from ``sloshing'' of the cluster core gas initiated by mergers. We present a set of simulations of galaxy cluster mergers with subclusters in order to determine the amount of heating provided by the mechanism of sloshing, exploring a parameter space over mass ratio, impact parameter, and viscosity of the intracluster medium (ICM). Our results show that for sloshing caused by mergers with gasless subclusters cooling may be partially offset by heating from sloshing, but this mechanism is less effective if the ICM is viscous.
\end{abstract}

\maketitle

\section{INTRODUCTION}

A long-standing open question with regard to clusters of galaxies is how to prevent overcooling of the gas in their cores. Though the most promising avenue for such a source of feedback appears to be outbursts from active galactic nuclei \cite[e.g.][]{dun06, mcn07}, an alternative method for heating the cores of galaxy clusters is that of central ``sloshing'' of the core gas. The presence of sloshing in the cores of relaxed galaxy clusters is inferred by the presence of ``cold fronts'' seen in {\it Chandra} and {\it XMM-Newton} observations \cite[e.g.][]{mar01, MVF03, MV07}. Hydrodynamic simulations \cite{tit05,asc06} have shown that such fronts can be caused by the sloshing of gas in the gravitational potential well caused by disturbances by passing subclusters. The process of sloshing brings hot, high-entropy gas at higher radii into contact with the cold, low-entropy gas of the core, resulting in the heating of the latter.

\section{THE SLOSHING PROCESS}

In principle any disturbance strong enough to perturb the cluster core significantly may cause sloshing, but the most likely candidate is a merger with a subcluster. As the subcluster approaches the main cluster's core and makes its first passage, the gas and dark matter peaks of the main cluster feel the same gravity force toward the subcluster and move together towards it. However, after the passage of the core, the direction of the gravitational force quickly changes, resulting in a change of sign of the ram pressure force on the gas peak. As a result of the compression from this ram pressure, the gas peak is pushed out of the potential minimum largely defined by the dark matter and out of hydrostatic equilibrium. As the cool gas from the core climbs out of the potential minimum, it reaches a maximum height and falls back into the potential well. It then overshoots the potential minimum and begins to climb out of the well on the other side. As it does this it expands and cools, forming a contact discontinuity (cold front). This process repeats itself, forming fronts of smaller amplitude. Throughout this process the lower entropy gas from the core is brought into contact with higher entropy gas from larger radii, and as these gases mix the entropy of the core is increased. 

\begin{figure}
  \includegraphics[height=.3\textheight]{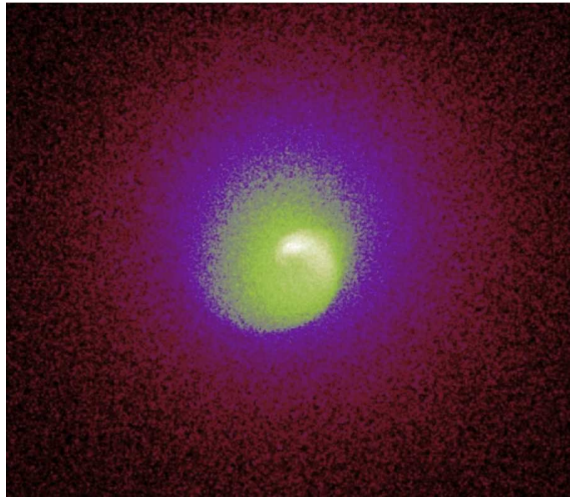}
  \caption{Simulated {\it Chandra} Image of core gas sloshing caused by the passage of a subcluster with a mass ratio $R$ = 5, and initial impact parameter $b$ = 500 kpc.}
\end{figure}

\section{FLASH SIMULATIONS OF CLUSTER-SUBCLUSTER MERGERS}

To investigate the possibility that the sloshing of the core gas in galaxy clusters may provide a significant form of feedback against radiative cooling, we have performed a set of binary cluster merger simulations using the \texttt{FLASH} code. We have explored a parameter space over initial impact parameter $b$ and initial mass ratio $R$, and considered scenarios where the ICM is modelled as an inviscid fluid and where there is an explicit viscosity present. The simulations presented here have initial mass ratios and impact parameters of $R$ = 5, $b$ = 500 kpc, and $R$ = 20, $b$ = 200 kpc; in each case a simulation was run with and without a constant kinematic viscosity approximately equal to the Spitzer value at a radius of $r$ = 50 kpc for the cluster. The simulations are adiabatic and do not include the effects of cooling.

\section{HEATING UP THE CORE}

Our simulations show the amount of heating provided by sloshing is dependent on two factors. Stronger encounters with subclusters will result in sloshing of higher radial amplitudes, bringing hotter gas into the cluster core and resulting in correspondingly higher heating rates. The degree of mixing also depends sensitively on the strength of the viscous forces that may be operating in the ICM. Viscosity damps motions in the ICM, suppressing instabilities (e.g., Kelvin-Helmholtz) and turbulence that would otherwise vigorously mix the cluster gas \cite{age07,mit09}. Therefore, it should be expected that in the presence of significant viscosity that the amount of heat provided by sloshing would less than if the ICM was inviscid.

\begin{figure}
  \includegraphics[height=.3\textheight]{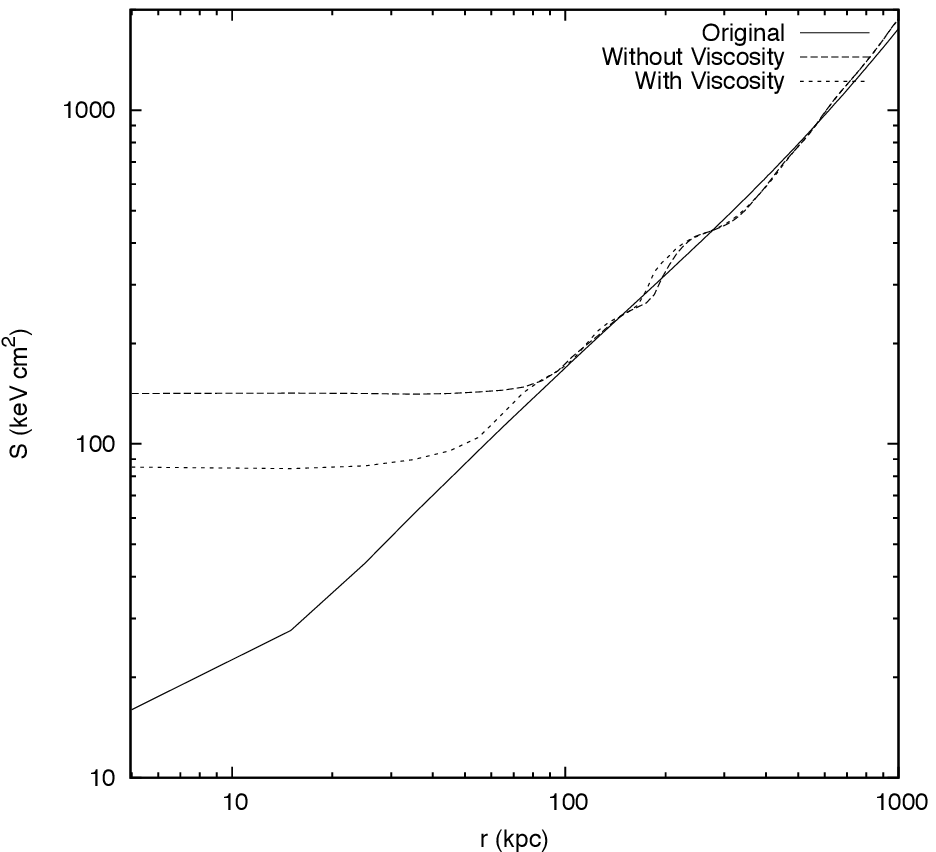}
  \includegraphics[height=.3\textheight]{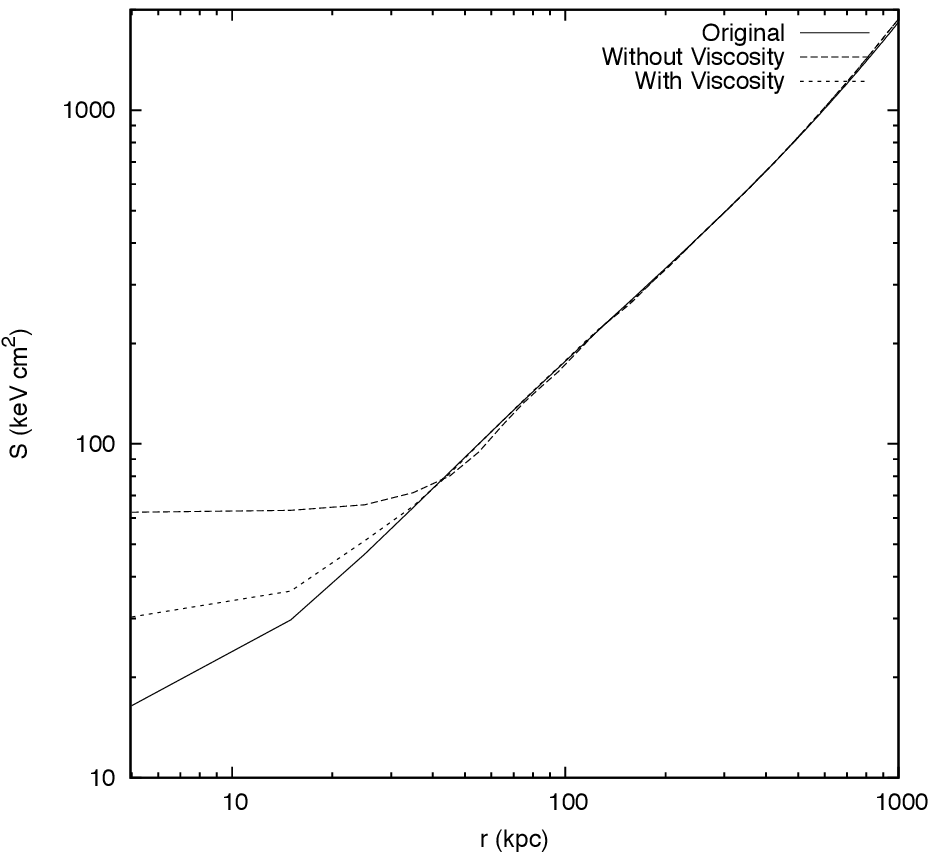}
 \caption{Entropy profiles at the end of the simulations ($t$ = 6.0 Gyr), compared to the original. Left: $R$ = 5, $b$ = 500 kpc. Right: $R$ = 20, $b$ = 200 kpc.\label{fig:entr}}
\end{figure}

\begin{figure}
  \includegraphics[height=.3\textheight]{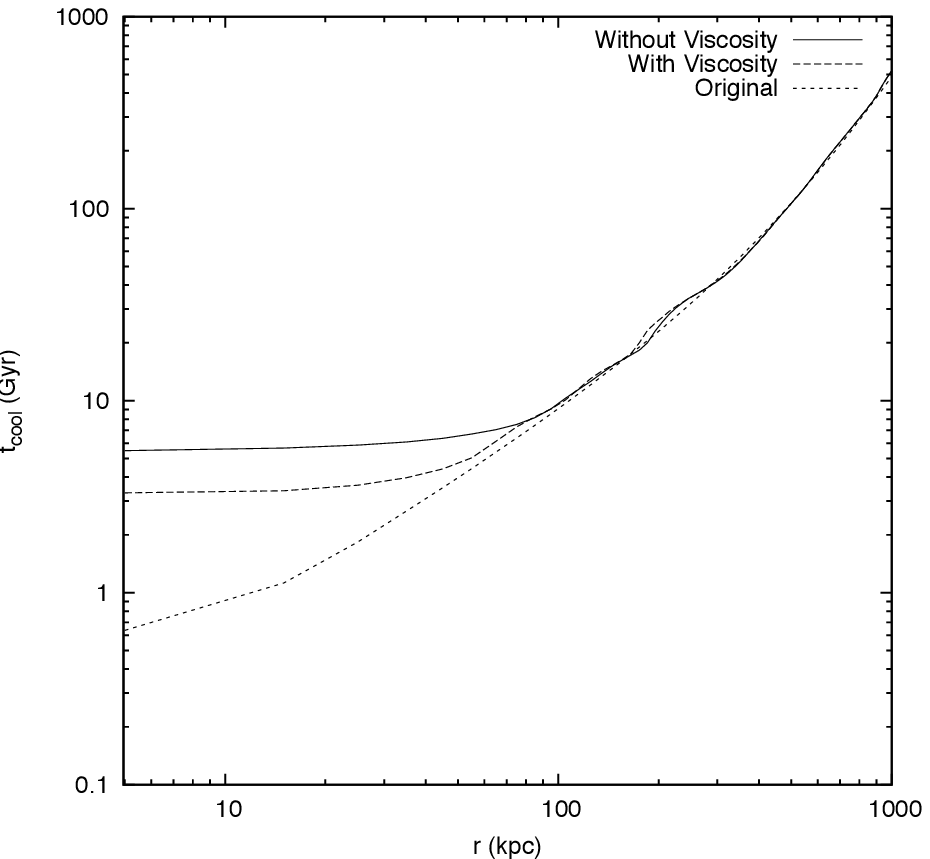}
  \includegraphics[height=.3\textheight]{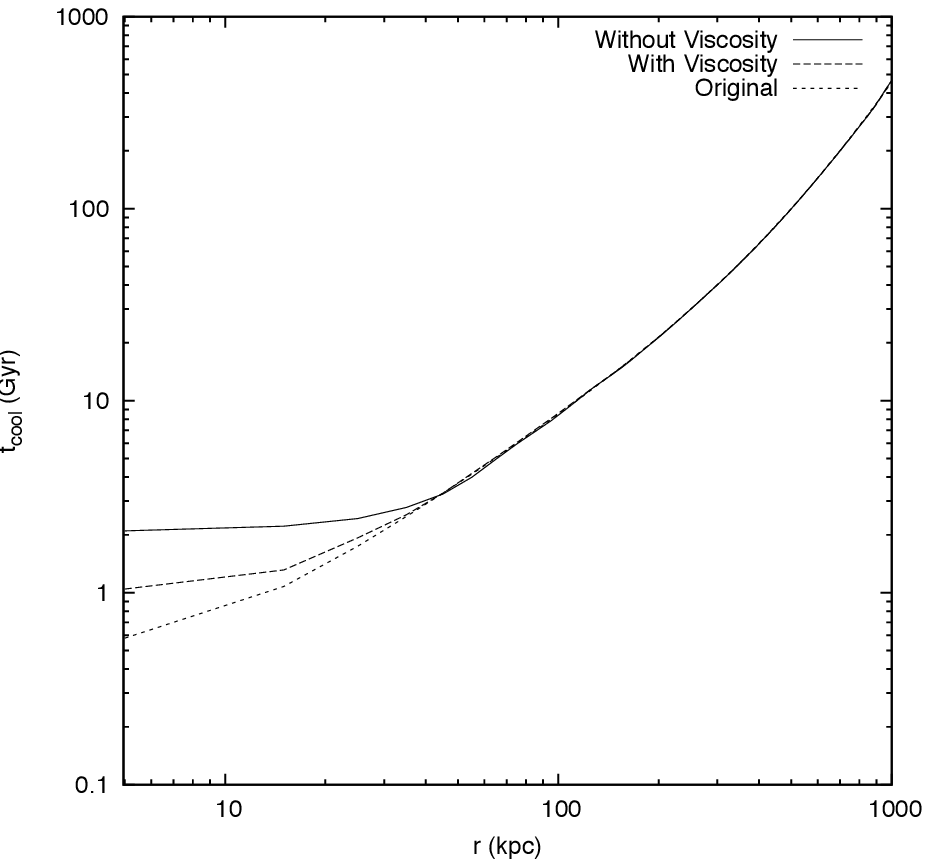}
 \caption{Cooling time profiles at the end of the simulations ($t$ = 6.0 Gyr), compared to the original. Left: $R$ = 5, $b$ = 500 kpc. Right: $R$ = 20, $b$ = 200 kpc.\label{fig:tcool}}
\end{figure}

Figure \ref{fig:entr} shows the final radial entropy profiles of the four simulations at an epoch $t$ = 6.0 Gyr after the beginning of the simulation, compared to the initial profile. In each case it is seen that the sloshing of the central gas has increased the entropy floor of the cluster core. Similarly, in Figure \ref{fig:tcool} the final radial cooling time profiles of the four simulations are plotted with the inital profile. The final central cooling times have increased from their initial values. Two important facts may be noted from these figures. The higher mass ratio encounters result in higher entropy floors and longer central cooling times. Additionally, it is seen that the degree of heating that results in the simulations in which the ICM is viscous is less than in the inviscid cases; the central entropies and cooling times are less by a factor of $\sim$1.5-2. 

Though the simulations do not include cooling explicitly, we may determine the effectiveness of the heating provided by sloshing with respect to the amount of cooling that would occur. Table \ref{tab:heat_to_cool} shows the ratio of the average heating rate to cooling rate for each simulation within three different radii from the cluster center. These results confirm what is seen in Figures \ref{fig:entr} and \ref{fig:tcool}: the heating rate is stronger with decreasing mass ratio and in the inviscid simulations. However, only for the $R$ = 5, $b$ = 500 kpc, inviscid case does the heating offset a significant amount of the cooling. At least in the simulations presented here sloshing may only provide a supplemental form of feedback.

\begin{table}
\begin{tabular}{lcccc}
\hline
  & \tablehead{1}{c}{b}{$R$ = 5\\$b$ = 500 kpc\\}
  & \tablehead{1}{c}{b}{$R$ = 5\\$b$ = 500 kpc\\w/ viscosity}
  & \tablehead{1}{c}{b}{$R$ = 20\\$b$ = 200 kpc\\}
  & \tablehead{1}{c}{b}{$R$ = 20\\$b$ = 200 kpc\\w/ viscosity}   \\
\hline
$r$ < 25 kpc & 76\% & 47\% & 19\% & 6\% \\
$r$ < 50 kpc & 60\% & 34\% & 8\% & 3\% \\
$r$ < 75 kpc & 45\% & 25\% & 4\% & 2\% \\
\hline
\end{tabular}
\caption{Ratio of average heating to average cooling in each of the simulations within different radii from the cluster center.}
\label{tab:heat_to_cool}
\end{table}

We plan to extend our investigation of the effect of the sloshing of cluster core gas on the heating of the ICM by expanding our parameter space over different mass ratios and impact parameters, and by simulating mergers with subclusters that include gas. Additionally, to determine accurately how long a runaway cooling flow can be mitigated by sloshing, we will need to perform simulations which self-consistently include the effects of cooling.


\begin{theacknowledgments}
JAZ was supported by {\it Chandra} grant GO8-9128X.
\end{theacknowledgments}

\end{document}